\documentclass[preprint]{emulateapj}
\usepackage{natbib}
\usepackage{graphicx}
\usepackage{amsmath}
\usepackage{empheq}
\allowdisplaybreaks

\begin{document}

\title{Medium-sized satellites of large Kuiper belt objects}

\author{Michael E. Brown} 
\affil{California Institute of Technology, Pasadena CA 91125 (U.S.A.)}

\author{Bryan J. Butler}
\affil{National Radio Astronomy Observatory, Socorro NM 87801 (U.S.A.)}

\begin{abstract}
While satellites of mid- to small-Kuiper belt objects tend to be
similar in size and brightness to their primaries, the largest 
Kuiper belt objects preferentially have satellites with small
fractional brightness. In the two cases where the sizes and albedos
of the small faint satellites have been measured, these satellites
are seen to be small icy fragments consistent with collisional
formation. Here we examine Dysnomia and Vanth, the
satellites of Eris and Orcus, respectively. Using the Atacama Large
Millimeter Array, we obtain the first spatially resolved 
observations of these systems at thermal wavelengths. Vanth is easily
seen in individual images and we find a 3.5$\sigma$ detection of Dysnomia by
stacking all of the data on the known position of the satellite. We calculate 
a diameter for Dysnomia of 700$\pm$115 km and for Vanth of 475$\pm$75 km, with albedos
of 0.04$^{+0.02}_{-0.01}$ and 0.08$\pm$0.02 respectively. Both Dysnomia and Vanth are indistinguishable
from typical Kuiper belt objects of their size. Potential implications
for the formation of these types of satellites are discussed.
\end{abstract}

\section{Introduction}
Most of largest objects in the Kuiper belt are known to have one or
more satellites orbiting the parent body. The majority of these satellites
have a small fractional brightness compared to their parent body.
Even before the discovery of any
of these small satellites, models predicted that giant impacts onto
differentiated bodies would preferentially form icy satellites with a 
small fractional mass \citep{2005Sci...307..546C}. Many of the known satellites to large 
Kuiper belt objects (KBOs) appear consistent with this paradigm. 
In
the two cases where compositional information
 of these small satellites is 
available, these satellite surfaces are known to have a high albedo
and to be dominated by water ice. The small satellites of Pluto
have been directly imaged by the New Horizons spacecraft and have
measured albedos of 0.5-0.9 and deep water ice absorptions in the near infrared
\citep{2016Sci...351.0030W,2017LPI....48.2478C}, 
while the satellites of Haumea show deep water ice absorption
\citep{2006ApJ...640L..87B,2009ApJ...695L...1F}, and dynamical modeling strongly suggests low
mass and thus high albedo \citep{2009AJ....137.4766R}.

Little is known about the size or albedo of
other satellites around large KBOs owing to the difficulty of
resolving the satellites at anything other than optical or
near-infrared wavelengths. The recently improved capability of
the {\it Atacama Large Millimeter Array} (ALMA) to obtain spatial
resolutions of 10s of milliarcseconds,
however, allows us to now measure thermal emission directly 
from KBO satellites.
Here, we use spatially resolved observations from ALMA 
to examine the size and albedo of two satellite systems:
Eris-Dysnomia and Orcus-Vanth.
Dysnomia, with a fractional brightness of 0.2\% that of Eris \citep{2007Sci...316.1585B}, 
appears to fit
the paradigm of small, icy,  collisionally-induced satellites surrounding
all of the largest known dwarf planets \citep{2006ApJ...639L..43B, 2016ApJ...825L...9P,
2017ApJ...838L...1K}.
A closer look
at the system, however, makes this assessment less certain. The unusually high
albedo of Eris of 0.97 \citep{2011_Sicardy} makes Dysnomia's relative brightness
seem artificially low. In fact, if Dysnomia has a typical small-KBO-like
albedo of $\sim$5\%, it is as large as 630 km. 
On the other  hand, if Dysnomia has an icy-collisional-satellite
like albedo of 0.5 or higher it is smaller than 200 km in radius. 
This range in sizes spans a wide
range of the types of satellite systems in the Kuiper belt. Without a
constraint on the size of Dysnomia, we lack a fundamental understanding
of this system.
A counter-example is the dwarf planet Orcus, which has a satellite -- Vanth --
with a fractional brightness of 9.6\% and a spectrum with significantly
less water ice than its primary \citep{2010AJ....139.2700B}. The origin of this
type of dwarf planet system remains uncertain, with models from 
capture to collision being plausible \citep{2009PhDT.......159R}.

\section{Observations}

Observations of Orcus-Vanth and Eris-Dysnomia were undertaken with the 12-m
array of the Atacama Large Millimeter Array (ALMA).  This synthesis
array is a collection of radio antennas, each 12 m in diameter, spread
out on the Altiplano in the high northern Chilean Andes.  Each of the
pairs of antennas acts as a two element interferometer, and the
combination of all of these individual interferometers allows for the
reconstruction of the full sky brightness distribution, in both
dimensions \citep{Thompson2001}.

ALMA is tunable in 7 discrete frequency bands, from $\sim$90 to $\sim$
950 GHz.  All observations in this paper were taken in Band 7, near 350
GHz, in the ``continuum'' (or ``TDM'') mode, with the standard frequency
tunings.  The data is observed in four spectral windows in this mode,
which for us had frequency ranges: 335.5--337.5 GHz; 337.5--339.5 GHz;
347.5--349.5 GHz; and 349.5--351.5 GHz.  In the final data analysis we
average over the entire frequency range in both bands, and use 345 GHz
as the effective frequency in our modeling.  All of these observations
are in dual-linear polarization; in the end we combine these into a
measurement of the total intensity.  

Table 1 shows the observational circumstances of our data.
 The Eris-Dysnomia system was observed in November and
December of 2015; The Orcus-Vanth system was observed in October and
November of 2016.  Initial calibration of the data was provided by the
ALMA observatory, and is done in the CASA reduction package via the ALMA
pipeline \citep{Muders2014}.  After the initial calibration, the data
product was a set of visibilities for each of the observing dates.

At this point we exported the data from CASA and continued the data reduction
in the AIPS package.  Because the primary purpose of the observations was to do
astrometry of the two systems, the observations were done in high-resolution
configurations of ALMA - with resolutions as fine as 15 mas.  For the
purposes of this paper, we are not concerned with such high resolution, but
rather simply enough resolution to distinguish the primary from the satellite.
Because of this we made images using weighting of the data which sacrifice
resolution for sensitivity (so-called "natural weighting").  The resulting
images are shown in Figures 1 and 2.

The final step of the data analysis was to estimate the observed flux density
for the primary and satellite for each observation.  For each image, we
obtained the values in a number of ways, to check for consistency: flux density
in the image; flux density in the CLEAN components; fitting a gaussian in the
image; and fitting the visibilities directly.  We found relatively good
agreement for all of these techniques.  We note that for the visibility fits,
we used point sources for all but Eris, where ALMA does slightly resolve the
body.  For Eris, we used a fit of a slightly limb-darkened disk, with radius of
1163 km \citep{2011_Sicardy}.  We take the visibility fit value as the best
value, as it avoids the biases of the image plane \citep{Greisen2004}.

There is one final correction that must be made to the flux densities; a
correction for atmospheric decorrelation.  For interferometric
observations, the Earth's atmosphere causes phase fluctuations in the
measured data which will result in a net reduction in the measured flux
density \citep{Thompson2001}.  In theory, and under good atmospheric
conditions, normal calibration will account for this decorrelation in
terms of the overall flux density scale, but image plane effects will
still persist (source broadening, for instance).  In normal ALMA
observations, decorrelation is only a minor effect, because observations
are scheduled when atmospheric conditions are good for the frequency
being observed.  However, our observations were done with specific
constraints -- namely that they had to be done in a particular (high
resolution) configuration, and that they had to be done with particular
time separations, in order to facilitate the astrometry.  Because of
these constraints, our observations were not done under optimal conditions in all
cases.  Fortunately, the way that astrometric observations are done with
ALMA provides a convenient method for correcting for the decorrelation.
Along with normal calibrations, astrometric "check sources" are
observed.  These check sources are point sources with well-known
astrometric positions.  Because they are observed with the same time
cadence as our target sources, and because they are relatively strong,
self-calibration \citep{1999_Cornwell} can be used to estimate how much
the flux density of these check sources changes from what the original
calibration indicates.  We used one of the check sources in each of the
observations to measure the magnitude of this effect, and applied it to
our final estimate of flux densities.  We note that this correction was
typically small for the Eris-Dysnomia observations, but much larger for
some of the Orcus-Vanth observations.  
Table 1 shows the final fitted flux densities, including all
corrections, for all of our observations.  

\begin{deluxetable*}{ccccc}
\caption{ \label{obs-tab} Observing dates, geometries, and flux-densities. }

\tablehead{
\colhead{bodies} & \colhead{date/time} & \colhead{Distance} & \colhead{primary f.d.} & \colhead{secondary f.d.} \\
\colhead{}       & \colhead{(UTC)}     & \colhead{(AU)}     & \colhead{(mJy)}        & \colhead{(mJy)} \\}
\startdata
Orcus-Vanth &  11 Oct 2016/11:02-12:16  & 48.8 & 1.160 $\pm$ 0.030 & 0.310 $\pm$ 0.030 \\
Orcus-Vanth &  13 Oct 2016/09:54-11:08  & 48.8 & 1.120 $\pm$ 0.060 & 0.270 $\pm$ 0.060 \\
Orcus-Vanth &  15 Oct 2016/11:54-12:59  & 48.7 & 1.180 $\pm$ 0.040 & 0.370 $\pm$ 0.040 \\
Orcus-Vanth & \phm{1}7 Nov 2016/09:30-10:40  & 48.4  & 1.170 $\pm$ 0.030 & 0.400 $\pm$ 0.030 \\
Eris-Dysnomia & 2015-Nov-09/03:25-04:35 & 95.4 & 0.803 $\pm$ 0.076 & --- \\
Eris-Dysnomia & 2015-Nov-13/02:50-04:00 & 95.4 & 0.893 $\pm$ 0.071 & --- \\
Eris-Dysnomia & 2015-Dec-04/01:15-02:25 & 95.7 & 0.825 $\pm$ 0.080 & --- \\
\enddata
\end{deluxetable*}

\begin{figure}
\plotone{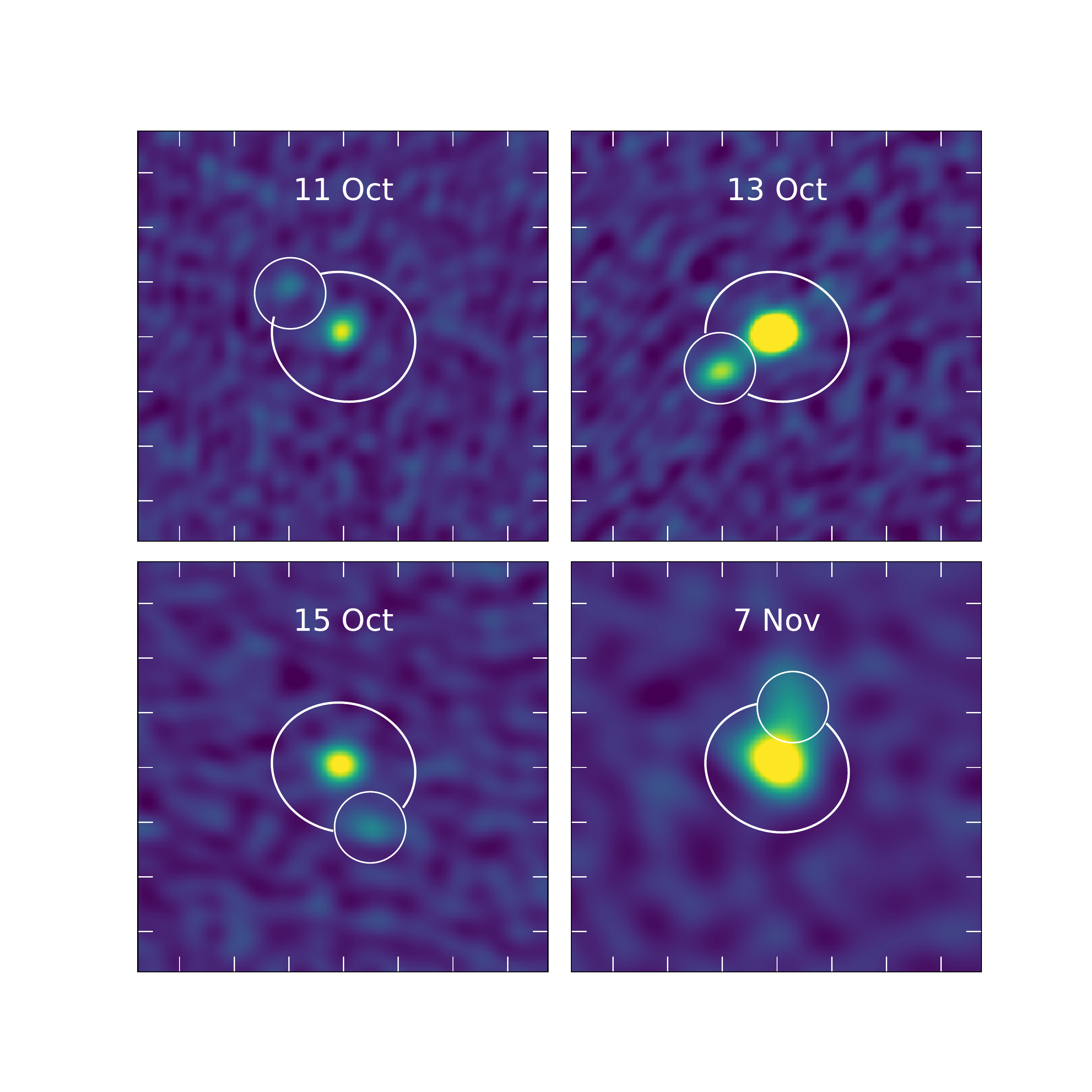}
\caption{ALMA observations of the Orcus-Vanth system. The images are
centered on Orcus and the predicted position of Vanth is circled. Tick 
marks in the images are 200 mas. Vanth is clearly detected even in
the lower resolution 7 Nov data.}
\end{figure}
\begin{figure}
\plotone{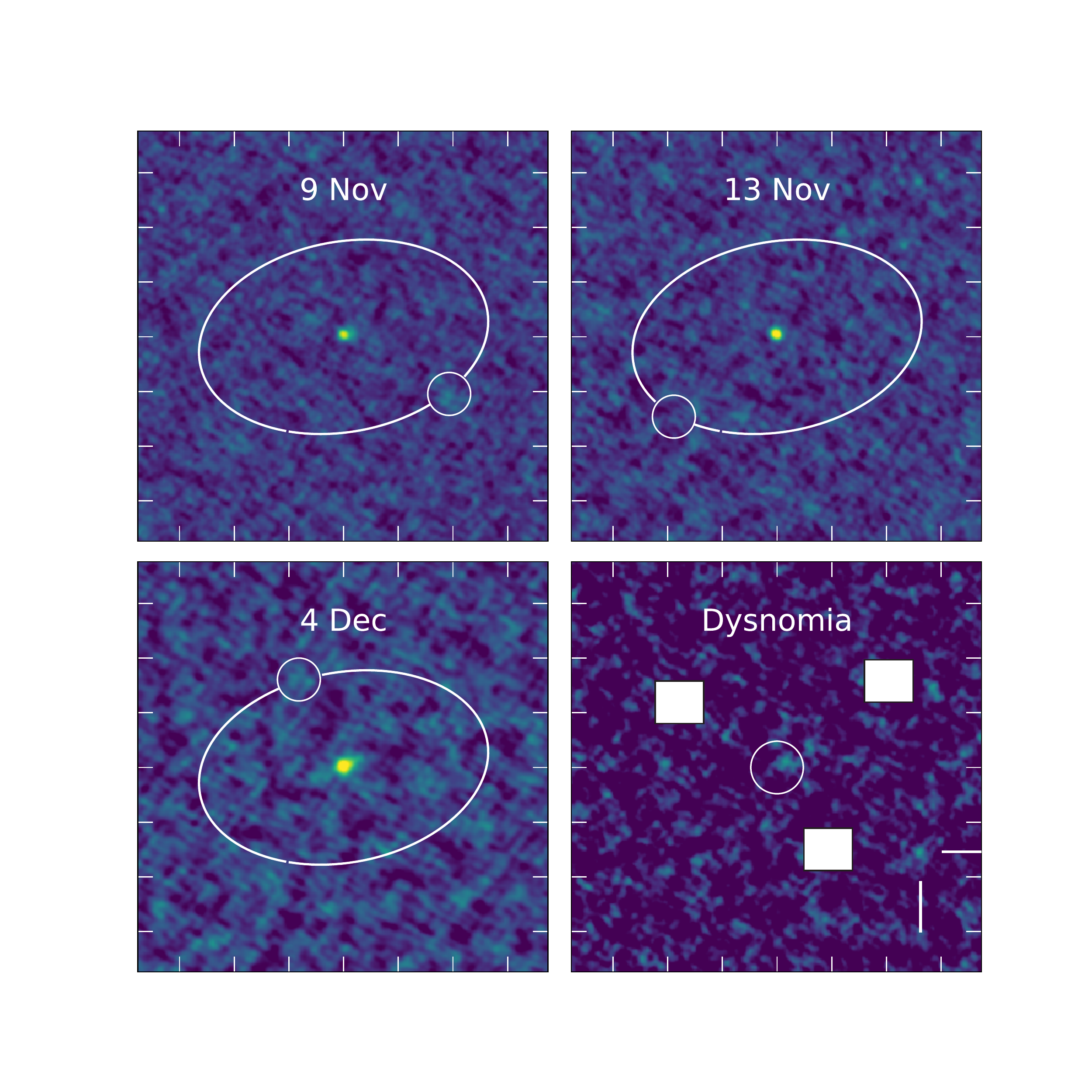}
\caption{ALMA observations of the Eris-Dysnomia system. The images are
centered on Eris and the predicted orbit of Dysnomia is show. Predicted
positions of Dysnomia based on contemporary HST observations are circled. 
In the bottom right panel, the three images are shifted and stacked 
at the position of Dysnomia, yielding a 3.5$\sigma$ detection of 
a source at the position of the satellite. The three positions of the shifted image
of Eris are masked with white boxes. In the full frame, only one other detection as
significant (marked with cross hairs in the lower right corner of the image) is seen.}
\end{figure}

\bigskip
\section{Orcus-Vanth}
A secondary source approximately 250 mas away from Orcus
is clearly visible in 3 of the 4 ALMA images, and a two-gaussian
fit also picks one out in the lower-resolution image from 2016 Nov 7. 
Using published Vanth orbital elements \citep{2010AJ....139.2700B,2011A&A...534A.115C} we
find that these detections are all along the orbital path of
Vanth and consistent with the predicted position if the mean
anomaly of Vanth is increased by 11 degrees, well within the
current uncertainties.
Flux densities measured for Orcus and Vanth are shown in
Table~1.

We use the measured thermal emission to determine the sizes of Orcus
and Vanth using the techniques detailed in \citet[][hereafter BB17]{2017AJ....154...19B}.
In our analysis we use a standard thermal model to calculate
the thermal emission expected from a distant body. In this model, the
free parameters are
bolometric emmisivity, phase integral, 
albedo, diameter, and a beaming parameter to account for the combined effects
of viewing geometry and surface thermal properties.
For the emissivity and phase integral,
 we use typical Kuiper belt object assumptions: 
we constrain emissivity to be between 0.8 and 1.0,
as argued in BB17, and we use the \citet{2009Icar..201..284B} 
empirical fit of 
phase integral to albedo with an allowed 50\% variation from these values.
Orcus and Vanth are not constrained to have any identical parameters.

We use the Markov Chain Monte Carlo scheme described in BB17 
to explore the best fit parameters
and their uncertainties. 
We fit the unresolved Orcus+Vanth
Spitzer 24 and 71 $\mu$m fluxes \citep{2008ssbn.book..161S},
the unresolved Herschel 70, 100, and 350 $\mu$m flux \citep{2013_Fornasier}, 
and
the new resolved ALMA data. We assume an 850 $\mu$m emissivity of
0.685, as derived in BB17, consistent with the value also found 
in \citet{2017A&A...608A..45L}. Figure 3 shows a collection of 30
random samples from the MCMC ensemble. The fit of the model to the
data is excellent.
\begin{figure}
\plotone{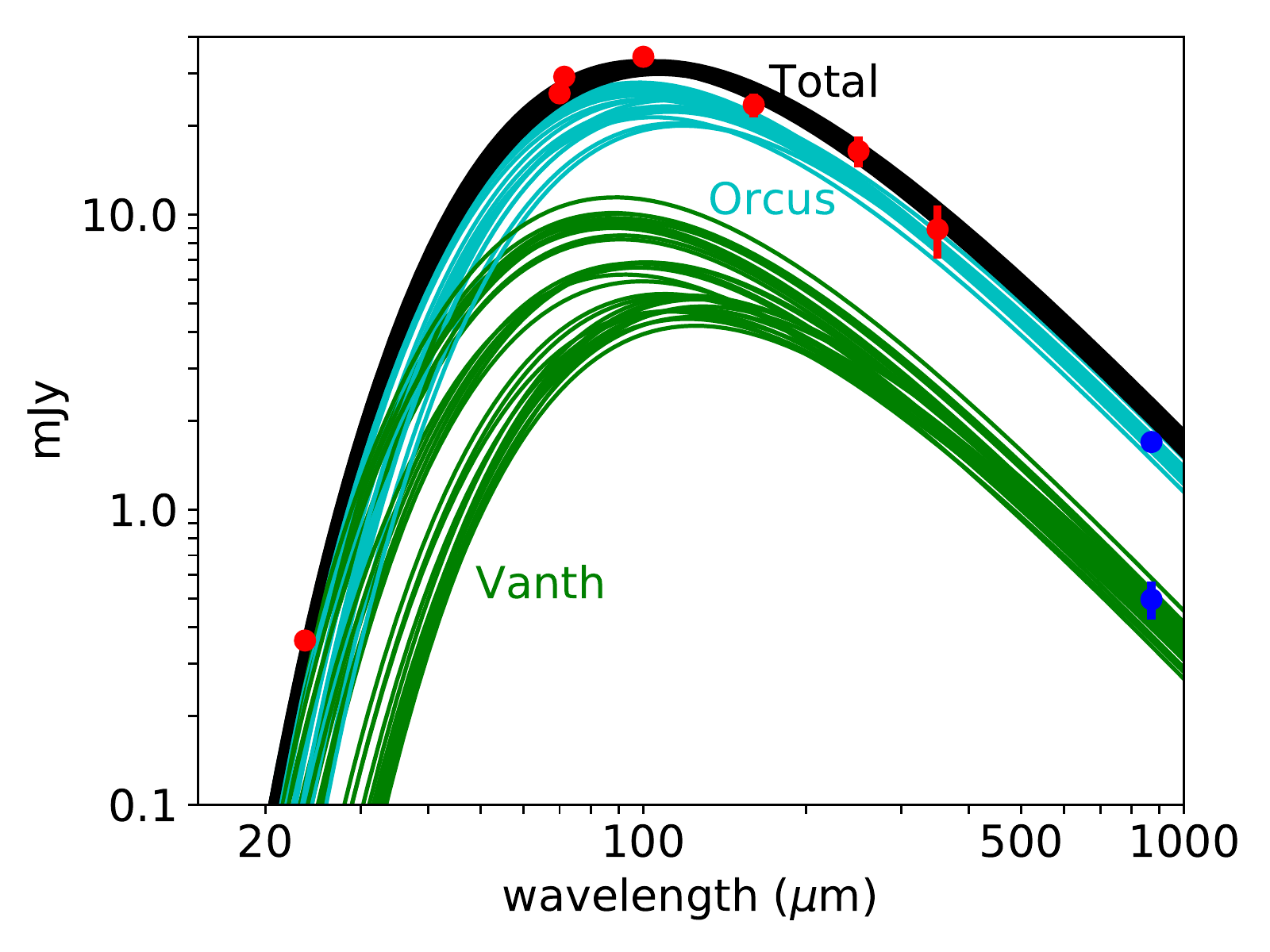}
\caption{A collection of 30 random samples from the MCMC ensemble compared to
data from the Orcus-Vanth system. The red points are unresolved data from
Spitzer and Herschel, while the blue points are the resolved data points from
ALMA. The ALMA data have been scaled by the inverse of the assumed emissivity
of 0.685 so that they appear at the equivalent emissivity of unity locations.}
\end{figure}


The marginalized distributions for the size and albedo of
both bodies are nearly gaussian, so we report the median and 
16th and 84th percentiles as our 1$\sigma$ error range. We find that
Orcus has a diameter of $910^{+50}_{-40}$ km and an albedo of
0.23$\pm$0.02, while Vanth has a diameter of 475$\pm$75 km
and an albedo of 0.08$\pm$0.02. Vanth is approximately
half of the diameter of Orcus, with an albedo approximately 3
times smaller. These results put Vanth within the range of typical
KBO albedos for objects of this size.

Without a knowledge of the density of Vanth, the mass ratio of the system
is unclear. For plausible densities from 0.8 g cm$^{-3}$ (the typical density 
for a $\sim$500 km object) up to 1.4 g cm$^{-3}$ (the system density if Orcus and
Vanth have identical densities)
the mass ratio ranges between 5 and 20, while the density of Orcus ranges 
from 1.4$\pm$0.2 to 2.0$\pm$0.3 g cm$^{-3}$.
Clearly, determining the density of Vanth is critical to understanding how
the Orcus-Vanth systems fits into our understanding of large KBOs and their
densities.

\section{Eris-Dysnomia}
No obvious detections of Dysnomia appear in the data. If we 
knew the predicted position of Dysnomia with respect to Eris,
however, we could make a more stringent determination.
The last published orbital elements of Dysnomia 
\citep{2007Sci...316.1585B} have a 30 degree phase uncertainty by the time
of these observations, so are not sufficient for providing 
predictions. We thus use archival observations obtained using
WF3 on the  Hubble Space Telescope in January 2015 to update the orbital
elements of Dysnomia and precisely predict its position
at the time of the ALMA observations 9 months later. Astrometric offsets
between Eris and Dysnomia for the times of observation
are determined using the methods detailed in \citet{2007Sci...316.1585B}.
Table 2 shows the relative positions of Dysnomia at the times of
the HST observations.

\begin{deluxetable}{cccc}
\tablecaption{Positions of Dysnomia}
\tablehead{
\colhead{JD} & \colhead{$\Delta$ RA} & \colhead{$\Delta$ Dec} & \colhead{observatory}\\
\colhead{} & \colhead{mas} & \colhead{mas} & \colhead{} }
\startdata
2457051.699   &  -347$\pm$2    & -226$\pm$1 & HST (measured)\\
2457054.950   &   \phm{-}282$\pm$3    &-325$\pm$2 & HST (measured)\\
\\
2457335.667   &  -383.2$\pm$1.4&-210.3$\pm$1.0 & ALMA (predicted)\\
2457339.641   &	  \phm{-}379.5$\pm$1.5&-291.5$\pm$1.7 &ALMA (predicted)\\
2457360.576   &	  \phm{-}159.8$\pm$1.7&\phm{-}322.8$\pm$1.0 &ALMA (predicted)\\
\enddata
\end{deluxetable}

\begin{deluxetable}{lll}
\tablecaption{Orbital elements of Dysnomia}
\tablehead{}
\startdata
semimajor axis&  \phm{15.}37460$\pm$80& km \\
inclination& \phm{5678}61.1$\pm$0.3 & deg \\
period& 15.78586$\pm$0.00008&  days \\
eccentricity& \phm{1234}$<.004$ \\
long. ascending node& \phm{678}139.6$\pm$0.2& deg \\
mean anomaly& \phm{678}273.2$\pm$0.02& deg \\
epoch (JD, defined)& \phm{123}2457054.95 \\
\enddata
\tablecomments{relative to J2000 ecliptic}
\end{deluxetable}

We calculate the new orbit of Dysnomia using the methods
described in \citet{2007Sci...316.1585B} with the
updated improvements using a Markov Chain Monte Carlo
scheme as described in \citet{2013ApJ...778L..34B}. 
The orbit
is in agreement with the previous results but with
improved uncertainties. The orbit continues to be consistent
with being circular, with a 1$\sigma$ upper limit to the eccentricity
of 0.004, so for the final fit we constrain the orbit to be
purely circular (Fig 4).
Updated orbital elements are provided in Table 3.
\begin{figure}
\plotone{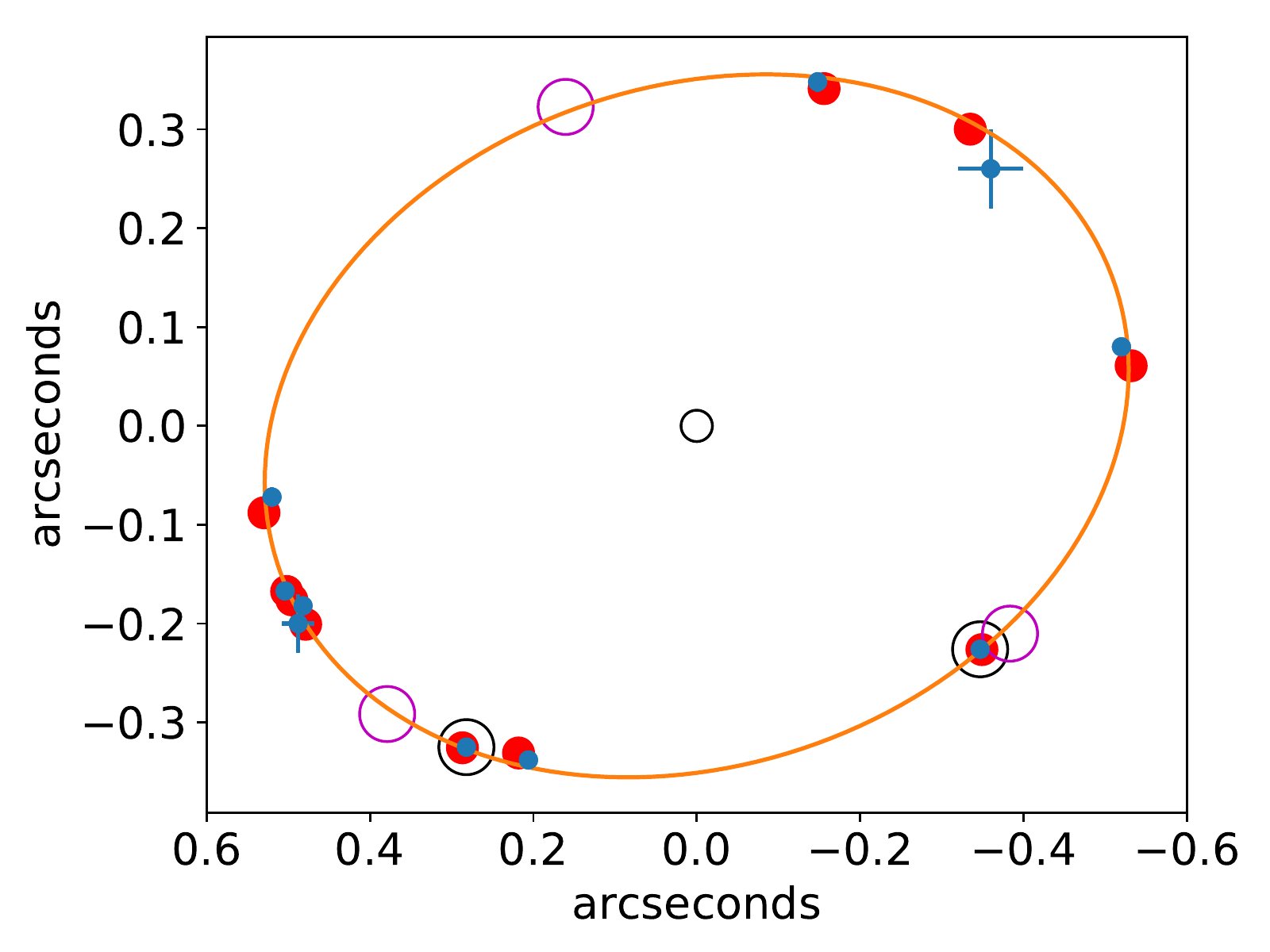}
\caption{A fit to the orbit of Dysnomia. Blue dots with error bars
are the observations, while their best-fit locations are show as red
dots. The two new data points added in this analysis are circled in 
blue. The large unfilled cyan circles are the predicted locations of
Dysnomia at the times of the ALMA observations.}
\end{figure}

The new orbit fit allows us to predict the locations of 
Dysnomia at the times of the ALMA observation nine months later
to within $\sim$2 mas, a fraction of the ALMA beam size.
Additional astrometric uncertainty  of order $\sim$10 mas ($1/4$ of the angular diameter
of Eris) arises because of the potential
offset between the center-of-light and center-of-mass of Eris
(if, for example, Eris has a warmer pole observed obliquely). 
We show these predicted locations in Figures 2 and 4. A flux density
enhancement occurs at the predicted location in at least
2 of the 3 observations. To increase the signal-to-noise of
a potential Dysnomia detection,
we shift all three images to be
centered on the predicted position of Dysnomia and sum them
(Figure 2). A source with a flux density of 0.13$\pm$0.03 mJy appears
16 mas from the expected position, smaller than the 30-50 mas resolution
element of the data and within the range of expected astrometric uncertainties
for such a low signal-to-noise source.

While we formally have a 4.2$\sigma$ detection of 
a potential source near the position of Dysnomia, 
we assess the true significance of this potential detection by 
performing an experiment where we shift each image by random
amounts and determine the maximum flux density within 30 mas
of this arbitrary 
position. We find values as high as the 0.13 mJy of Dysnomia only 0.02\% 
of the time, corresponding to a 3.5$\sigma$ likelihood that the detection
is indeed Dysnomia. As an additional check, we reexamine the combined image 
and find that in the 2.75 square arcsecond field
we only see one other source as strong as the potential Dysnomia detection. 
Our field consists of approximately 1400 
resolution elements, so we should expect to randomly detect a 3.5$\sigma$ source 33\%
of time, consistent with the observation. While the detection of Dysnomia is weak,
we can find no reason to discount its reality.
We proceed on the assumption that we have indeed 
detected Dysnomia.

We model the emission from Eris and Dysnomia using an identical technique
as previously used for Orcus-Vanth.
We assume the occultation-derived circular diameter of Eris  
of 2326$\pm$12 km \citep{2011_Sicardy} and have as the free parameters in our model fitting
the beaming parameter of Eris and the diameter, albedo, and
beaming parameter of Dysnomia. We again constrain the bolometric emissivity 
and Bond albedo as above. 

We first investigate if there is any evidence for emission from
Dysnomia from previous, unresolved, data. We use data at 71$\mu$m from the Spitzer
Space Telescope \citep{2008ssbn.book..161S} and data at 70, 100, and
150$\mu$m from the Herschel Space Telescope \citep{2012A&A...541A..92S}.
The marginalized distribution for the diameter of Dysnomia
gives a 1$\sigma$ upper limit of 810 km and 
a corresponding 1$\sigma$ lower limit for albedo of 0.03.
The unresolved data provide essentially no information on the 
parameters of Dysnomia. Nonetheless, if no Dysnomia is included in
the fit, the flux of the shortest wavelength data
is under-predicted, suggesting the possibility that a large dark body
could be present in the system (Figure 5).
\begin{figure}
\plotone{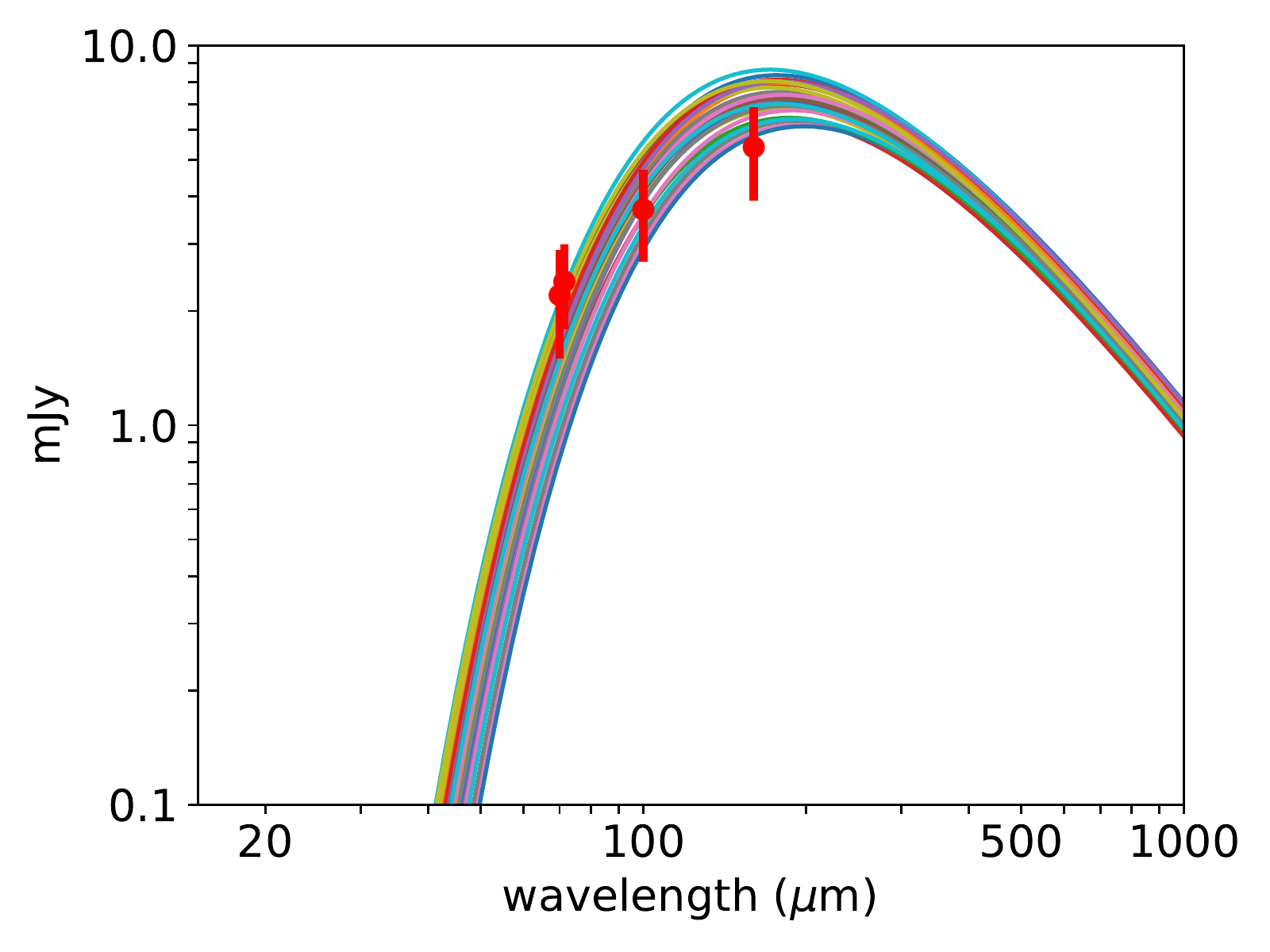}
\caption{A collection of 30 random samples from the MCMC ensemble compared to
data from the Eris-Dysnomia system. We model only a single body and
the unresolved measurements from Herschel. While the fit does not statistically
support the existence of a second body in the system, it is clear that
a smaller darker (thus warmer) body could improve the fit to the data.}
\end{figure}

We next fit the resolved ALMA data together with the previous
unresolved data. We continue to use our earlier-derived 850 $\mu$m
emissivity of 0.685 for Dysnomia, assuming it is a typical KBO. 
For Eris, which has a significantly different surface composition,
we consider ALMA observations  of Pluto
\citep{2015DPS....4721004B}, which suggest significantly
lower brightness temperature, but it is difficult to separate the
effects of emissivity and cold atmospherically buffered N$_2$ ice. 
As N$_2$ ice at the temperature of Eris has significantly lower volatility,
we assume that ice effects are negligible and instead use the same canonical 0.685 value
for emissivity.
Figure 6 shows 30 samples from the MCMC ensemble to the thermal data. 
We find that Dysnomia has a diameter of 700$\pm$115 km with an albedo of 0.04$^{+0.02}_{-0.01}$.
Dysnomia's size and albedo are consistent with those of typical mid-sized KBOs, but the
albedo is nearly 25 times smaller than the extremely bright Eris. Allowing the
density of Dysnomia to range from as low as 0.8 g cm$^{-3}$ to as high as being equal to
that of Eris gives a range of system mass ratios between 37 and 115, again a large possible
range.

\begin{figure}
\plotone{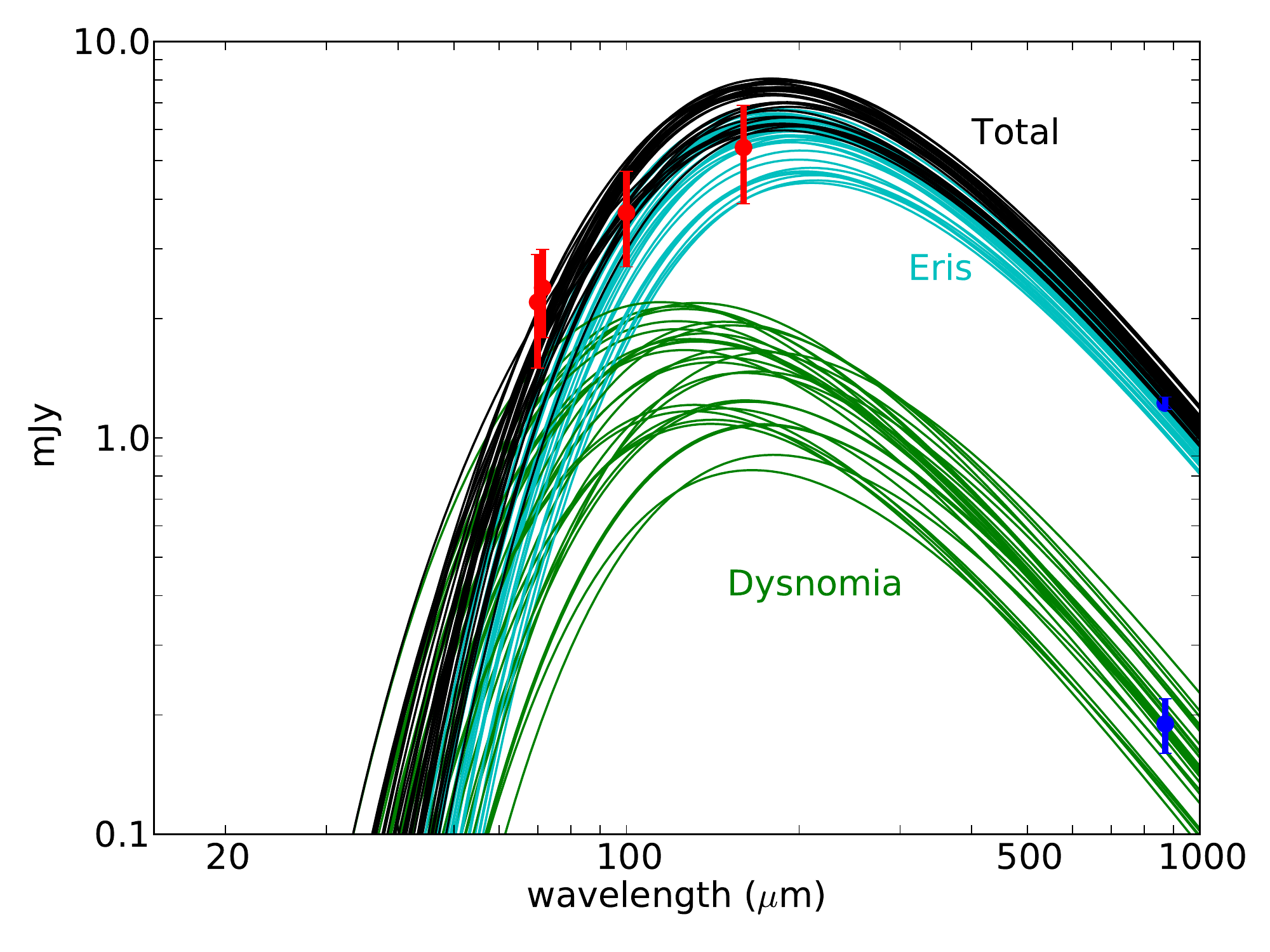}
\caption{A collection of 30 random samples from the MCMC ensemble compared to
data from the Eris-Dysnomia system. Including the resolved ALMA data
shows the presence of a mid-sized dark Dysnomia which
somewhat improves the fit to the shorter wavelength data where Dysnomia's contribution is 
more significant. }
\end{figure}

\section{Discussion}
Neither Vanth, the satellite of Orcus, nor Dysnomia, the satellite of Eris,
fits the paradigm of the type of
small icy collisionally derived satellite predicted by the
models of \citet{2005Sci...307..546C}. Such satellites form out of the icy disk surrounding the 
parent in the aftermath of the collision and would be expected to be nearly pure water ice like
the satellites of Haumea and the small satellites of Pluto appear to be. Indeed,
both Dysnomia and Vanth have the low albedos expected for typical KBOs of
their size \citep{2008ssbn.book..161S}. 
We consider possible formation mechanisms below.

Smaller multiple-KBO systems tend to be binaries with similar-sized components
\citep{2008ssbn.book..345N}. These are thought to either be formed through 
dynamical-friction-assisted capture \citep{2002Natur.420..643G} or to have
initially formed as a pair \citep{2010AJ....140..785N}. 
Little exploration has been done of the range of
possible satellite parameters that could ensue from such mechanisms,
but they are generally thought to be relevant to small KBO pairs. There
is no reason to expect that while mid-sized KBOs have a modest satellite
fraction, the largest KBOs would preferentially have
satellites owing to these same mechanisms.

Charon, the large satellite of Pluto, appears to have formed from an
grazing giant impact which then left the proto-Charon largely
intact, but with a low enough energy to remain bound \citep{2005Sci...307..546C}.
In simulations, collisions of undifferentiated bodies mostly
yielded a satellite whose composition was unchanged from the initial
impactor. The Orcus-Vanth system is plausibly explained by such a
scenario. With a mass ratio between 5 and 20,
Orcus-Vanth appears to be a good candidate for an
analog to the Pluto-Charon system (with a mass ratio of 8), 
with the exception that we detect no small icy analogs of the small Pluto satellite.
 \citet{2008ssbn.book..335B} places a
brightness ratio limit of 0.1\% for any more distant undiscovered
objects in the Orcus system. For an icy albedo of 0.5-1.0, this
limit corresponds to objects with a diameter of 15 to 20 km, smaller than
at least two of the small satellites of Pluto.

The Eris-Dysnomia system is unlike any other known in the Kuiper belt. 
With a mass ratio between 37 and 115, Dysnomia appears intermediate
between satellites such as Charon and Nyx and Hydra \citep[with mass ratios
greater than $10^5$;][]{2015Icar..246..317B}
Hi'iaka, the larger satellite of Haumea \citep[with a mass ratio of $\sim$200;][]{2009AJ....137.4766R}.
But an albedo of $\sim$0.04 clearly shows that
Dysnomia is not a reassembled product of an icy disk. We suggest two
alternatives. First, it is possible that our reported detection of
Dysnomia is erroneous. While we showed that the probability of a 
spurious detection at the
predicted combined location of Dysnomia is low, robust detection
at multiple locations are desired. Second, if the detection
of Dysnomia is real, as the statistics suggest, our understanding of
Kuiper belt satellite formation mechanisms is clearly inadequate and
the range of possible satellite formation outcomes is larger than 
currently thought. 

While large KBOs have a high fraction of satellites, a general understanding
of the diverse possible formation mechanisms for these satellites is lacking.
It is generally believed that giant impacts are responsible
for the satellites of these KBOs, but
simulations have only attempted to explain individual systems \citep{2005Sci...307..546C,
2011AJ....141...35C,2010ApJ...714.1789L},
and no comprehensive study has explored a wide range of possible outcomes.
\citet{2016MNRAS.460.1542B} have suggested a general paradigm where collisions
are either of the Charon-forming type or of the icy-small-fragment type, but it
is not clear if this paradigm can be reconciled with the implication that Dysnomia
appears not to have formed from a post-impact icy disk of material. 
As understanding of these satellite systems likely provides insights into
populations and collisions in the early outer solar system, emphasis should 
be placed on both the theoretical and observational exploration of 
these objects.

This  paper  makes  use  of  the  following  ALMA  data: ADS/JAO.ALMA\#2016.1.00830.S, ADS/JAO.ALMA\#2015.1.00810.S.  ALMA  is  a  partnership   of   ESO   (representing   its   member   states), NSF  (USA)  and  NINS  (Japan),  together  with  NRC (Canada), MOST and ASIAA (Taiwan), and KASI (Republic  of  Korea),  in cooperation  with  the  Republic  of Chile.   The  Joint  ALMA  Observatory  is  operated  by ESO, AUI/NRAO and NAOJ. The National Radio Astronomy  Observatory  is  a  facility  of  the  National  Science Foundation operated under cooperative agreement by Associated Universities, Inc.
by Associated Universities, Inc.


\begin{thebibliography}{34}
\expandafter\ifx\csname natexlab\endcsname\relax\def\natexlab#1{#1}\fi

\bibitem[{{Barkume} {et~al.}(2006){Barkume}, {Brown}, \&
  {Schaller}}]{2006ApJ...640L..87B}
{Barkume}, K.~M., {Brown}, M.~E., \& {Schaller}, E.~L. 2006, \apjl, 640, L87

\bibitem[{{Barr} \& {Schwamb}(2016)}]{2016MNRAS.460.1542B}
{Barr}, A.~C. \& {Schwamb}, M.~E. 2016, \mnras, 460, 1542

\bibitem[{{Brown}(2008)}]{2008ssbn.book..335B}
{Brown}, M.~E. 2008, in The Solar System Beyond Neptune, ed. M.~A. {Barucci},
  H.~{Boehnhardt}, D.~P. {Cruikshank}, \& A.~{Morbidelli}, 335--344

\bibitem[{{Brown}(2013)}]{2013ApJ...778L..34B}
---. 2013, \apjl, 778, L34

\bibitem[{{Brown} \& {Butler}(2017)}]{2017AJ....154...19B}
{Brown}, M.~E. \& {Butler}, B.~J. 2017, \aj, 154, 19

\bibitem[{{Brown} {et~al.}(2010){Brown}, {Ragozzine}, {Stansberry}, \&
  {Fraser}}]{2010AJ....139.2700B}
{Brown}, M.~E., {Ragozzine}, D., {Stansberry}, J., \& {Fraser}, W.~C. 2010,
  \aj, 139, 2700

\bibitem[{{Brown} \& {Schaller}(2007)}]{2007Sci...316.1585B}
{Brown}, M.~E. \& {Schaller}, E.~L. 2007, Science, 316, 1585

\bibitem[{{Brown} {et~al.}(2006){Brown}, {van Dam}, {Bouchez}, {Le Mignant},
  {Campbell}, {Chin}, {Conrad}, {Hartman}, {Johansson}, {Lafon}, {Rabinowitz},
  {Stomski}, {Summers}, {Trujillo}, \& {Wizinowich}}]{2006ApJ...639L..43B}
{Brown}, M.~E., {van Dam}, M.~A., {Bouchez}, A.~H., {Le Mignant}, D.,
  {Campbell}, R.~D., {Chin}, J.~C.~Y., {Conrad}, A., {Hartman}, S.~K.,
  {Johansson}, E.~M., {Lafon}, R.~E., {Rabinowitz}, D.~L., {Stomski}, Jr.,
  P.~J., {Summers}, D.~M., {Trujillo}, C.~A., \& {Wizinowich}, P.~L. 2006,
  \apjl, 639, L43

\bibitem[{{Brozovi{\'c}} {et~al.}(2015){Brozovi{\'c}}, {Showalter}, {Jacobson},
  \& {Buie}}]{2015Icar..246..317B}
{Brozovi{\'c}}, M., {Showalter}, M.~R., {Jacobson}, R.~A., \& {Buie}, M.~W.
  2015, \icarus, 246, 317

\bibitem[{{Brucker} {et~al.}(2009){Brucker}, {Grundy}, {Stansberry}, {Spencer},
  {Sheppard}, {Chiang}, \& {Buie}}]{2009Icar..201..284B}
{Brucker}, M.~J., {Grundy}, W.~M., {Stansberry}, J.~A., {Spencer}, J.~R.,
  {Sheppard}, S.~S., {Chiang}, E.~I., \& {Buie}, M.~W. 2009, Icarus, 201, 284

\bibitem[{{Butler} {et~al.}(2015){Butler}, {Gurwell}, {Lellouch}, {Moullet},
  {Moreno}, {Bockelee-Morvan}, {Biver}, {Fouchet}, {Lis}, {Stern}, {Young},
  {Young}, {Weaver}, {Boissier}, \& {Stansberry}}]{2015DPS....4721004B}
{Butler}, B.~J., {Gurwell}, M., {Lellouch}, E., {Moullet}, A., {Moreno}, R.,
  {Bockelee-Morvan}, D., {Biver}, N., {Fouchet}, T., {Lis}, D., {Stern}, A.,
  {Young}, L., {Young}, E., {Weaver}, H., {Boissier}, J., \& {Stansberry}, J.
  2015, in AAS/Division for Planetary Sciences Meeting Abstracts, Vol.~47,
  AAS/Division for Planetary Sciences Meeting Abstracts, 210.04

\bibitem[{{Canup}(2005)}]{2005Sci...307..546C}
{Canup}, R.~M. 2005, Science, 307, 546

\bibitem[{{Canup}(2011)}]{2011AJ....141...35C}
---. 2011, \aj, 141, 35

\bibitem[{{Carry} {et~al.}(2011){Carry}, {Hestroffer}, {DeMeo}, {Thirouin},
  {Berthier}, {Lacerda}, {Sicardy}, {Doressoundiram}, {Dumas}, {Farrelly}, \&
  {M{\"u}ller}}]{2011A&A...534A.115C}
{Carry}, B., {Hestroffer}, D., {DeMeo}, F.~E., {Thirouin}, A., {Berthier}, J.,
  {Lacerda}, P., {Sicardy}, B., {Doressoundiram}, A., {Dumas}, C., {Farrelly},
  D., \& {M{\"u}ller}, T.~G. 2011, \aap, 534, A115

\bibitem[{{Cook} {et~al.}(2017){Cook}, {Binzel}, {Cruikshank}, {Dalle Ore},
  {Earle}, {Ennico}, {Grundy}, {Howett}, {Jennings}, {Lunsford}, {Olkin},
  {Parker}, {Philippe}, {Protopapa}, {Reuter}, {Schmitt}, {Stansberry},
  {Stern}, {Verbiscer}, {Weaver}, {Young}, {New Horizons Composition Theme
  Team}, \& {Ralph Instrument Team}}]{2017LPI....48.2478C}
{Cook}, J.~C., {Binzel}, R.~P., {Cruikshank}, D.~P., {Dalle Ore}, C.~M.,
  {Earle}, A., {Ennico}, K., {Grundy}, W.~M., {Howett}, C., {Jennings}, D.~J.,
  {Lunsford}, A.~W., {Olkin}, C.~B., {Parker}, A.~H., {Philippe}, S.,
  {Protopapa}, S., {Reuter}, D., {Schmitt}, B., {Stansberry}, J.~A., {Stern},
  S.~A., {Verbiscer}, A., {Weaver}, H.~A., {Young}, L.~A., {New Horizons
  Composition Theme Team}, \& {Ralph Instrument Team}. 2017, in Lunar and
  Planetary Inst.~Technical Report, Vol.~48, Lunar and Planetary Science
  Conference, 2478

\bibitem[{{Cornwell} \& {Fomalont}(1999)}]{1999_Cornwell}
{Cornwell}, T. \& {Fomalont}, E.~B. 1999, in Astronomical Society of the
  Pacific Conference Series, Vol. 180, Synthesis Imaging in Radio Astronomy II,
  ed. G.~B. {Taylor}, C.~L. {Carilli}, \& R.~A. {Perley}, 187--199

\bibitem[{Fornasier {et~al.}(2013)Fornasier, Lellouch, M{\"u}ller, Santos-Sanz,
  Panuzzo, Kiss, Lim, Mommert, Bockelée-Morvan, Vilenius, Stansberry, Tozzi,
  Mottola, Delsanti, Crovisier, Duffard, Henry, Lacerda, Barucci, \&
  Gicquel}]{2013_Fornasier}
Fornasier, S., Lellouch, E., M{\"u}ller, T., Santos-Sanz, P., Panuzzo, P.,
  Kiss, C., Lim, T., Mommert, M., Bockelée-Morvan, D., Vilenius, E.,
  Stansberry, J., Tozzi, G.~P., Mottola, S., Delsanti, A., Crovisier, J.,
  Duffard, R., Henry, F., Lacerda, P., Barucci, A., \& Gicquel, A. 2013,
  Astronomy and Astrophysics, 555, id.A15

\bibitem[{{Fraser} \& {Brown}(2009)}]{2009ApJ...695L...1F}
{Fraser}, W.~C. \& {Brown}, M.~E. 2009, \apjl, 695, L1

\bibitem[{{Goldreich} {et~al.}(2002){Goldreich}, {Lithwick}, \&
  {Sari}}]{2002Natur.420..643G}
{Goldreich}, P., {Lithwick}, Y., \& {Sari}, R. 2002, \nat, 420, 643

\bibitem[{Greisen(2004)}]{Greisen2004}
Greisen, E.~W. 2004, in Astronomical Society of the Pacific Conference Series,
  Vol. 281, Astronomical Data Analysis Software and Systems XI, ed. D.~A.
  Bohlender, D.~Durand, \& T.~H. Handley, 260--264

\bibitem[{{Kiss} {et~al.}(2017){Kiss}, {Marton}, {Farkas-Tak{\'a}cs},
  {Stansberry}, {M{\"u}ller}, {Vink{\'o}}, {Balog}, {Ortiz}, \&
  {P{\'a}l}}]{2017ApJ...838L...1K}
{Kiss}, C., {Marton}, G., {Farkas-Tak{\'a}cs}, A., {Stansberry}, J.,
  {M{\"u}ller}, T., {Vink{\'o}}, J., {Balog}, Z., {Ortiz}, J.-L., \& {P{\'a}l},
  A. 2017, \apjl, 838, L1

\bibitem[{{Leinhardt} {et~al.}(2010){Leinhardt}, {Marcus}, \&
  {Stewart}}]{2010ApJ...714.1789L}
{Leinhardt}, Z.~M., {Marcus}, R.~A., \& {Stewart}, S.~T. 2010, \apj, 714, 1789

\bibitem[{{Lellouch} {et~al.}(2017){Lellouch}, {Moreno}, {M{\"u}ller},
  {Fornasier}, {Santos-Sanz}, {Moullet}, {Gurwell}, {Stansberry}, {Leiva},
  {Sicardy}, {Butler}, \& {Boissier}}]{2017A&A...608A..45L}
{Lellouch}, E., {Moreno}, R., {M{\"u}ller}, T., {Fornasier}, S., {Santos-Sanz},
  P., {Moullet}, A., {Gurwell}, M., {Stansberry}, J., {Leiva}, R., {Sicardy},
  B., {Butler}, B., \& {Boissier}, J. 2017, \aap, 608, A45

\bibitem[{Muders {et~al.}(2014)Muders, Wyrowski, Lightfoot, Williams, Nakazato,
  Kosugi, Davis, \& Kern}]{Muders2014}
Muders, D., Wyrowski, F., Lightfoot, J., Williams, S., Nakazato, T., Kosugi,
  G., Davis, L., \& Kern, J. 2014, in Astronomical Data Analysis Software and
  Systems XXIII, ed. N.~Manset \& P.~Forshay, 383--386

\bibitem[{{Nesvorn{\'y}} {et~al.}(2010){Nesvorn{\'y}}, {Youdin}, \&
  {Richardson}}]{2010AJ....140..785N}
{Nesvorn{\'y}}, D., {Youdin}, A.~N., \& {Richardson}, D.~C. 2010, \aj, 140, 785

\bibitem[{{Noll} {et~al.}(2008){Noll}, {Grundy}, {Chiang}, {Margot}, \&
  {Kern}}]{2008ssbn.book..345N}
{Noll}, K.~S., {Grundy}, W.~M., {Chiang}, E.~I., {Margot}, J.-L., \& {Kern},
  S.~D. 2008, in The Solar System Beyond Neptune, ed. M.~A. {Barucci},
  H.~{Boehnhardt}, D.~P. {Cruikshank}, \& A.~{Morbidelli}, 345--363

\bibitem[{{Parker} {et~al.}(2016){Parker}, {Buie}, {Grundy}, \&
  {Noll}}]{2016ApJ...825L...9P}
{Parker}, A.~H., {Buie}, M.~W., {Grundy}, W.~M., \& {Noll}, K.~S. 2016, \apjl,
  825, L9

\bibitem[{{Ragozzine} \& {Brown}(2009)}]{2009AJ....137.4766R}
{Ragozzine}, D. \& {Brown}, M.~E. 2009, \aj, 137, 4766

\bibitem[{{Ragozzine}(2009)}]{2009PhDT.......159R}
{Ragozzine}, D.~A. 2009, PhD thesis

\bibitem[{{Santos-Sanz} {et~al.}(2012){Santos-Sanz}, {Lellouch}, {Fornasier},
  {Kiss}, {Pal}, {M{\"u}ller}, {Vilenius}, {Stansberry}, {Mommert}, {Delsanti},
  {Mueller}, {Peixinho}, {Henry}, {Ortiz}, {Thirouin}, {Protopapa}, {Duffard},
  {Szalai}, {Lim}, {Ejeta}, {Hartogh}, {Harris}, \&
  {Rengel}}]{2012A&A...541A..92S}
{Santos-Sanz}, P., {Lellouch}, E., {Fornasier}, S., {Kiss}, C., {Pal}, A.,
  {M{\"u}ller}, T.~G., {Vilenius}, E., {Stansberry}, J., {Mommert}, M.,
  {Delsanti}, A., {Mueller}, M., {Peixinho}, N., {Henry}, F., {Ortiz}, J.~L.,
  {Thirouin}, A., {Protopapa}, S., {Duffard}, R., {Szalai}, N., {Lim}, T.,
  {Ejeta}, C., {Hartogh}, P., {Harris}, A.~W., \& {Rengel}, M. 2012, \aap, 541,
  A92

\bibitem[{Sicardy(2011)}]{2011_Sicardy}
Sicardy, B. . .~o. 2011, Nature, 478, 493

\bibitem[{{Stansberry} {et~al.}(2008){Stansberry}, {Grundy}, {Brown},
  {Cruikshank}, {Spencer}, {Trilling}, \& {Margot}}]{2008ssbn.book..161S}
{Stansberry}, J., {Grundy}, W., {Brown}, M., {Cruikshank}, D., {Spencer}, J.,
  {Trilling}, D., \& {Margot}, J.-L. 2008, in The Solar System Beyond Neptune,
  ed. M.~A. {Barucci}, H.~{Boehnhardt}, D.~P. {Cruikshank}, \& A.~{Morbidelli},
  161--179

\bibitem[{Thompson {et~al.}(2001)Thompson, Moran, \& Swenson}]{Thompson2001}
Thompson, A.~R., Moran, J.~M., \& Swenson, G.~W. 2001, Interferometry and
  Synthesis in Radio Astronomy, 2nd Edition (New York, New York:
  Wiley-Interscience)

\bibitem[{{Weaver} {et~al.}(2016){Weaver}, {Buie}, {Buratti}, {Grundy},
  {Lauer}, {Olkin}, {Parker}, {Porter}, {Showalter}, {Spencer}, {Stern},
  {Verbiscer}, {McKinnon}, {Moore}, {Robbins}, {Schenk}, {Singer}, {Barnouin},
  {Cheng}, {Ernst}, {Lisse}, {Jennings}, {Lunsford}, {Reuter}, {Hamilton},
  {Kaufmann}, {Ennico}, {Young}, {Beyer}, {Binzel}, {Bray}, {Chaikin}, {Cook},
  {Cruikshank}, {Dalle Ore}, {Earle}, {Gladstone}, {Howett}, {Linscott},
  {Nimmo}, {Parker}, {Philippe}, {Protopapa}, {Reitsema}, {Schmitt}, {Stryk},
  {Summers}, {Tsang}, {Throop}, {White}, \& {Zangari}}]{2016Sci...351.0030W}
{Weaver}, H.~A., {Buie}, M.~W., {Buratti}, B.~J., {Grundy}, W.~M., {Lauer},
  T.~R., {Olkin}, C.~B., {Parker}, A.~H., {Porter}, S.~B., {Showalter}, M.~R.,
  {Spencer}, J.~R., {Stern}, S.~A., {Verbiscer}, A.~J., {McKinnon}, W.~B.,
  {Moore}, J.~M., {Robbins}, S.~J., {Schenk}, P., {Singer}, K.~N., {Barnouin},
  O.~S., {Cheng}, A.~F., {Ernst}, C.~M., {Lisse}, C.~M., {Jennings}, D.~E.,
  {Lunsford}, A.~W., {Reuter}, D.~C., {Hamilton}, D.~P., {Kaufmann}, D.~E.,
  {Ennico}, K., {Young}, L.~A., {Beyer}, R.~A., {Binzel}, R.~P., {Bray}, V.~J.,
  {Chaikin}, A.~L., {Cook}, J.~C., {Cruikshank}, D.~P., {Dalle Ore}, C.~M.,
  {Earle}, A.~M., {Gladstone}, G.~R., {Howett}, C.~J.~A., {Linscott}, I.~R.,
  {Nimmo}, F., {Parker}, J.~W., {Philippe}, S., {Protopapa}, S., {Reitsema},
  H.~J., {Schmitt}, B., {Stryk}, T., {Summers}, M.~E., {Tsang}, C.~C.~C.,
  {Throop}, H.~H.~B., {White}, O.~L., \& {Zangari}, A.~M. 2016, Science, 351,
  aae0030

\end{thebibliography}
\end{document}